\documentclass[fleqn,10pt]{wlscirep}
\usepackage[utf8]{inputenc}
\usepackage[T1]{fontenc}
\usepackage{lineno}

\usepackage{amsmath} 
\usepackage{amsthm} 
\usepackage{amssymb} 
\usepackage{epstopdf}
\usepackage{enumerate}
\usepackage{multicol} 
\usepackage{wrapfig}
\usepackage{color}
\usepackage{tensind}
\tensordelimiter{`}
\whenindex{a}{\alpha}{} \whenindex{b}{\beta}{} \whenindex{g}{\gamma}{} \whenindex{d}{\delta}{} \whenindex{e}{\varepsilon}{} \whenindex{m}{\mu}{} \whenindex{n}{\nu}{} \whenindex{l}{\lambda}{} \whenindex{k}{\kappa}{} \whenindex{r}{\rho}{} \whenindex{s}{\sigma}{} \whenindex{t}{\tau}{}
\usepackage{todonotes}
\usepackage{subcaption}
\usepackage{setspace}
\usepackage{comment}
\usepackage{cancel}
\usepackage[]{units}
\usepackage{hyperref}
\usepackage[all]{hypcap}
\makeatletter 
\usepackage[bb=boondox]{mathalfa} 
\renewcommand{\v}[1]{\ensuremath{\vec{#1}}} 

\let\baraccent=\= 
\renewcommand{\=}[1]{\stackrel{#1}{=}} 

\renewcommand{\[}{\left [}
\renewcommand{\]}{\right ]}
\newcommand{\<}{\left <}
\renewcommand{\>}{\right >}

\newtheorem{cnj}{Conjecture}

\theoremstyle{definition}

\theoremstyle{remark}

\newtheorem{?}{\textbf{Question}}
\renewcommand{\exp}[1]{\mbox{exp}\[#1\]} 

\title{Emergent order from mixed chaos at low temperature}

\author[1,*]{Pavel Chvykov}
\author[2]{Jeremy England}
\affil[1]{Institute for Globally Distributed Open Research and Education (IGDORE)}
\affil[2]{Visiting Professor, Department of Physics, Bar-Ilan University, Israel}

\affil[*]{pchvykov@gmail.com}

\begin{abstract}
This paper explores a novel connection between a thermodynamic and a dynamical systems perspective on emergent dynamical order. 
We provide evidence for a conjecture that Hamiltonian systems with mixed chaos spontaneously find regular behavior when minimally coupled to a thermal bath at sufficiently low temperature. Numerical evidence across a diverse set of five dynamical systems supports this conjecture, and allows us to quantify corollaries about the organization timescales and disruption of order at higher temperatures. 
Balancing the damping-induced phase-space contraction against thermal exploration, we are able to predict the transition temperatures in terms of the relaxation timescales, indicating a novel nonequilibrium fluctuation-dissipation relation, and formally connecting the thermodynamic and dynamical systems views.
Our findings suggest that for a wide range of real-world systems, coupling to a cold thermal bath leads to emergence of robust, non-trivial dynamical order, rather than a mere reduction of motion as in equilibrium. 
\end{abstract}
\begin{document}

\flushbottom
\maketitle

\thispagestyle{empty}
\section{Introduction}

Emergence of order in space, time, or response properties in dynamical systems generally depends on the selection of a low-entropy or low-phase-space-volume ensemble of microstates, which is stabilized by system dynamics relative to a much wider range of microstates that were available in principle.  Thus, understanding general mechanisms by which physical dynamics can lead to the selection of fine-tuned patterns has broad relevance to the study of self-organization and emergent order in nature and applications \cite{Chvykov_Berrueta2021, jin2024metamat}.

Prigogine is known for pointing out that dissipation of energy from an external drive seems to play a key role in various examples of self-organization\cite{goldbeter2018dissipative}. In particular, he established the intuition that certain kinds of non-trivial dynamical or spatial order could only emerge once the strength of external driving pushed the system in question beyond the linear-response regime\cite{glansdorff1973thermodynamic}. However, Prigogine's program ultimately failed to yield a general theory of the so-called dissipative structures he was credited with characterizing\cite{cross1993pattern}.

An interesting hint indicating how Prigogine may have been right comes from the more mathematical work in dynamical systems theory, where there is a common intuition that adding weak damping can lead to emergence of order in mixed-chaotic systems\cite{lieberman1985transient, jousseph2016weak, Elyutin_2004a, coccolo2024fractional}. It is, however, unclear how generally or formally this intuition is thought to apply. Additionally, the focus of past work on this phenomenon has been in low-dimensional systems, making the connection to many-body self-organization tenuous. Still, insofar as damping in physical systems is a mechanism of dissipation, there may as a result turn out to be a broad class of  dynamical systems that might be said to exhibit Prigogine's dissipative structure.

The key to putting this connection on firmer physical ground is to realize that damping and temperature are inextricably linked. Subjecting a Hamiltonian system to damping without noise, as was typically done in the aforementioned mixed chaos studies, is equivalent to coupling the system to a zero temperature thermal bath. The Einstein relation $b\,D=k_{B} T$, where $D$ is the diffusion constant, $b$ is damping coefficient, and $T$ is temperature, quantifies this connection, which arises because the bath of particles responsible for the damping drag must also be the source of random thermal fluctuations. 

In this sense, the order arising from noiseless damping in mixed-chaotic dynamical systems could potentially be attributed to the impact of being at low temperature. The low-temperature regime is known to bring about spontaneous order in thermal equilibrium, but can it produce a similar effect for driven dynamical systems?  In this paper, we study the phenomenon of emergent order brought on by coupling mixed-chaotic dynamical systems to a low-temperature thermal bath. We thus combine the perspectives of dynamical systems theory (via damping) and thermodynamics (via low temperature), and argue for the generality of this mechanism, including in many-body systems.

The clearest thermodynamic argument for emergent order comes from cases involving equilibrium self-assembly, in which a multitude of attractive and repulsive inter-particle forces in a many-body system on the microscale can lead to the formation of useful meso- and macrostructures such as proteins and crystals \cite{pelesko2007self_ass, MCMANUS2016protein_self_ass}. In typical cases of equilibrium self-assembly, there is a clear explanation for the low-entropy phenomenon that comes from the effect of temperature: when such systems decrease in average energy, they are no longer free to explore a vast diversity of collective states and instead become trapped in rare local energy minima. 


Away from equilibrium, for a dynamical system subject to time-varying external forces, we can write a generalization of the Boltzmann distribution for the steady-state of a system with microstates $x_i$ at energies $u_i$ and temperature $k_B T$:
\begin{equation} \label{eq:neqW}
    p(x_{i}) \propto \exp{-\frac{u_i}{k_{B}T}} \; \bigg
    \langle\exp{-\frac{w_{\rightarrow i}}{k_{B}T}}\bigg\rangle ^{-1}
\end{equation}
The path-dependence characteristic of nonequilibrium systems is captured here by the second term, which depends on the work $w_{\to i}$ done by external driving on the system as it relaxes from an arbitrary initial starting state into the final state $i$. The averaging $\<\cdot\>$ here is done over all stochastic relaxation paths (allowing that the notation for $w_{\to i} $ implies taking the required $t\rightarrow \infty$ limit only after normalizing $p$).
Considering the case of constant $u_i$ in order to focus on purely dynamical effects, we can see that just as in equilibrium self-assembly, lowering temperature has the potential to concentrate the probability in a rare set of states, resisting the disordering pull of entropy. Here, however, such states are selected for their exceptionally positive-leaning distribution of dissipative work history $w_{\to i}$, instead of for their low energy. But what kind of order might these states and associated trajectories reflect? And how is that order stabilized without the attractive interparticle forces that hold things together in the equilibrium case? \cite{Marsland_England_2017, england2015dissipative}.


The dynamical systems theory perspective on mixed chaotic systems helps to provide an explanation.
Mixed chaos is defined as the coexistence of both chaotic and regular (non-chaotic) behaviors in different regions of the dynamical system’s phase space. While this might seem unusual at first, it turns out that this is the ``rule rather than the exception'' among non-linear Hamiltonian dynamical systems \cite{percival1987chaos}. Intuitively, this happens because it is less likely for a real-world system to be 100\% chaotic or regular, than for it to have some of both in its phase space. Formally, this arises as a consequence of KAM (Kolmogorov–Arnold–Moser) theory, which tells us that mixed chaos dynamics will usually emerge when we add small nonlinear interactions to an otherwise linear (or integrable) dynamical system. As the nonlinearity is increased, progressively larger regions of the phase space can become chaotic, while regular behavior can be maintained in some corners of the phase space -- leaving islands of regularity in the sea of chaos (fig.\ref{fig:4sys}-iii).

The emergent order in dynamical systems theory is understood in terms of damping, rather than low temperature as in thermodynamics\cite{Hu_Wang_Hu_Zhou_Liu_2020, Androulidakis_Alexandridis_Konstantopoulos_2013, Elyutin_2004b} . The key effect of damping on a Hamiltonian system is to break the phase-space volume conservation of Liouville's theorem.  For undamped dynamics
$\dot{p} = -\partial H/\partial q$ and $\dot{q}= \partial H/\partial p $, the identity $\nabla\cdot[\dot{q},\dot{p}]=\partial^{2}H/\partial q\partial p-\partial^{2}H/\partial p\partial q=0$ ensures incompressible flow in phase space ($d\rho/dt = 0$) so that each infinitesimal element $dq~dp$ cannot be contracted or expanded over time. Thus, a non-dissipative Hamiltonian system cannot lose memory of initial conditions. Additionally, chaotic and ordered dynamical regions of phase space are forbidden to intermix since the closed regular orbits inside the islands of regularity cannot suddenly start moving chaotically, either forward or backward in time. By contrast, introduction of damping via $\dot{p} = -\partial H/\partial q - bp$, results in $d\rho/dt =-\rho\nabla\cdot[\dot{q},\dot{p}] = b\rho $ , which permits phase-space volume contraction into a low-entropy subset of states. Damping is necessary for such concentration of probability, but not sufficient, as chaos in principle could stretch phase-space volumes into space-filling filaments. In mixed chaotic systems, however, initially chaotic trajectories would be expected to explore diffusively until coming arbitrarily close to an ordered trajectory, whereas ordered trajectories stay ordered over time. Due to this asymmetry, the accumulation of probability density in ordered regions becomes favored. Put another way, islands of regularity in Hamiltonian dynamics are regions with vanishing local Lyapunov exponents $\lambda_i=0$, meaning that volumes do not get stretched in any dimension.  The addition of weak damping drives exponents below zero and can make these regions attracting \cite{jousseph2016weak, jousseph2018hierarchical}. 

In this paper we combine thermodynamic and dynamical systems perspectives by developing numerical evidence and physical intuition for the following conjecture: 
\begin{cnj}
Hamiltonian dynamics exhibiting mixed chaos will settle into islands of order upon weakly coupling to a thermal bath at sufficiently low temperature. \label{conj:1}
\end{cnj}
In sec. \ref{sec:ss}, we demonstrate this phenomenon in a variety of example systems, including a many-body example suggesting relevance for self-organization. 
Any formal proofs of this conjecture are left to future work.
To better characterize the thermodynamic nature of the phenomenon, we further numerically study the relation between damping strength and relaxation time, discovering a power-law relation (sec. \ref{sec:time}), as well as the phase diagram that shows how order disappears at higher temperatures (sec. \ref{sec:T}). Relating these two, we show preliminary evidence for a possible nonequilibrium fluctuation-dissipation relation in these systems. 

\section{Steady-state} \label{sec:ss}
\begin{figure*}
\includegraphics[width=1\textwidth]{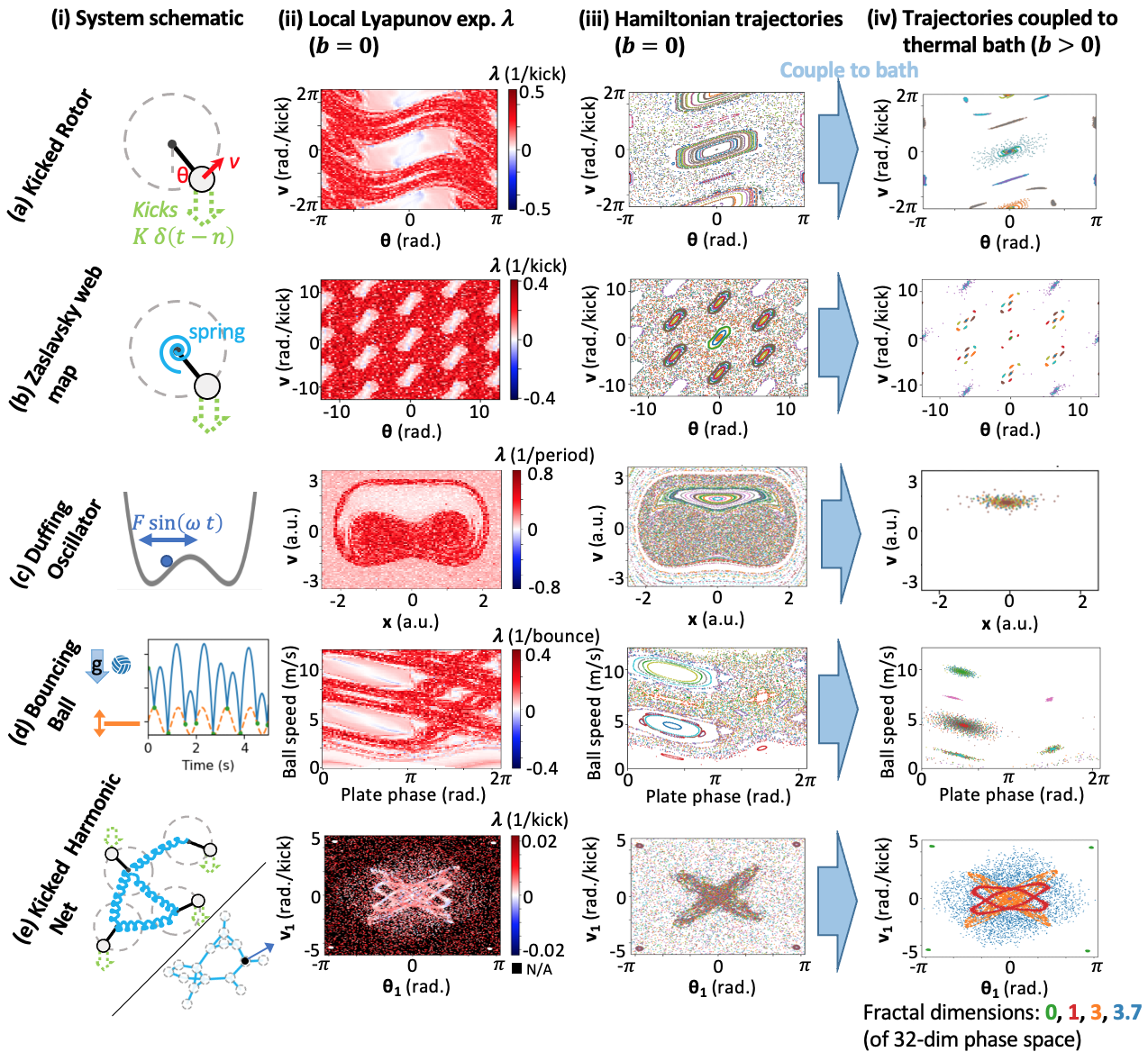}
\caption{Numerical evidence for conjecture \ref{conj:1} across 5 dynamical systems. Column (ii) shows the values of local Lyapunov exponents for points sampled throughout the system's phase, quantitatively identifying islands of regularity. These are then also seen in the system trajectories under baseline Hamiltonian dynamics in column (iii), which we then weakly couple to a thermal bath (add damping and noise) to get the steady-state dynamics pictured in column (iv). In (iii) and (iv), we plot stroboscopic system configurations 
for 256 initial conditions (IC), picked randomly in the shown phase space. 
Each IC and subsequent trajectory is plotted in its own color (so points of one color correspond to one trajectory). Each IC is evolved under one unique noise realization. The coordinates shown are: (a,b) rotor angle and speed right before kick, (e) same, but only for one of the 16 coupled rotors (the one marked black in the network in (e-i), (c) oscillator position and speed when $\omega\, t =0 $ (mod $2\pi$), (d) plate phase and ball speed immediately after each bounce. (e) is the only many-body example, and is more challenging to visualize: regular behavior is very rare in the 32-dim system phase space, and so none of the randomly sampled IC demonstrate it. The IC shown combine random sampling with sampling from the damped steady-state attractors. Also, since here partial organization is possible, attractors need not be 0-dimensional -- so we measure their fractal dimension to quantify degree of organization. \newline
In all cases, we see that bath coupling collapses all trajectories to the Lyapunov $\lambda = 0$ islands of regularity, giving emergent order.
}
\label{fig:4sys}
\end{figure*}

We begin by showing that our conjecture \ref{conj:1} works consistently across five different dynamical systems in-silico, as shown in the corresponding five rows (a-e) of fig.\ref{fig:4sys}. In each case, we see that regions that used to be islands of regularity under Hamiltonian dynamics (columns (ii) and (iii)), become global attractors as soon as we couple to the thermal bath at low temperature $T$ (column (iv)). Mathematically, by ``coupling to a thermal bath" we mean adding damping and noise forces to our system $\delta F = -b\,v + \sqrt{2\,b\,T}\;\xi(t) $, where $b$ is the damping coefficient and $ \xi(t) $ is white Gaussian noise process (such that $ \<\xi(t)\>=0 $ and $ \<\xi(t),\, \xi(t')\> = \delta(t-t') $). The noise amplitude is determined in terms of $ T $ via the Einstein relation. We can thus think of $b$ as the ``bath coupling strength,'' as it controls both terms, restoring Hamiltonian dynamics in the $b \to 0$ limit. This setup clarifies that adding damping without noise -- a simplification used in most literature on damped chaos -- corresponds to the idiosyncratic $T=0$ thermal bath. Also, throughout this work we assume the bath coupling to be a relatively weak perturbation on top of baseline Hamiltonian dynamics, meaning that it does not significantly change the local phase-space structure as heavy damping would\cite{jousseph2018hierarchical}. If damping was large, above a critical value it is known that new strange attractors can appear that have no signature in the undamped Hamiltonian system \cite{lieberman1985transient}.

To quantitatively characterize the islands of regularity, we measure the local Lyapunov exponents $\lambda_i$ throughout each system's phase space -- fig.\ref{fig:4sys} column (ii). To do this, for each point we start two trajectories at nearby initial conditions, and run them for 20 steps of our stroboscopic Poincare map (for 50 in (e)), monitoring how their separation grows. This helps to understand the role of local Lyapunov exponent in selecting the dynamically ordered behaviors under thermal bath coupling.

Remarkably, the five examples in fig.\ref{fig:4sys} show that our conjecture continues to hold regardless of how weak the bath coupling $ b $ is, and regardless of the degree of fine-tuning required (see fig.\ref{fig:time}) or of the complexity of the regular motion (e.g., see the attractor shapes in fig.\ref{fig:4sys}(e-iv). 

The first system we considered is the kicked rotor (or Chirikov standard map, fig.\ref{fig:4sys}a), which is the paradigmatic example of mixed chaos, and has been extensively studied in context of dynamical systems theory (as well as quantum chaos and Anderson localization) \cite{zaslavsky1992weak, delande2013kicked}. The standard setup is as a conservative Hamiltonian dynamics, which we then couple to a thermal bath (see \cite{Elyutin_2004a} for a more technical analysis of this system):
\begin{align}
\dot{\theta} &= v \nonumber\\
\dot{v} &= -K\,\sin(\theta)\;\sum_n\delta(t-n) - b\,v + \sqrt{2\,b\,T}\;\xi(t) \label{eq:KR}
\end{align}
This describes the dynamics $ \theta(t) $ of a particle hinged on a freely-rotating rigid rod, and periodically kicked by a global uniform force field $ K $ (where $ \delta $ is the Dirac delta and $ n $ -- any integer). The advantage of this system is that it's very fast to simulate, as we can integrate the linear dynamics between kicks analytically, effectively turning the system into a discrete stroboscopic map with a single simulation step per kick. The stochastic noise can similarly be added to the linear evolution by scaling its amplitude by $\sqrt{\Delta t}$ -- integration time. It has previously been noted that this system regularizes its motion under addition of damping $ b >0$ \cite{Elyutin_2004a, Martins_Gallas_2008}, and we will explore and extend this result.

$ K $ is the only parameter controlling the Hamiltonian system dynamics, and thus the phase-space structure. For $ K\lesssim 0.97 $, the entire phase-space is regular, resembling that of a simple pendulum, while for $ K\gtrsim 6.75 $ it is entirely chaotic. Between these two values, chaotic and regular regions coexist, and trajectories never cross from one to the other. This drastically changes as soon as we add any amount of damping $ b>0 $, which now causes all trajectories to collapse to islands of regularity in the steady-state (after sufficient time). Further addition of thermal noise $ T>0 $ will diffuse the steady-state density around these islands, much like around energy wells at thermodynamic equilibrium (for fig.\ref{fig:4sys}a, we used $ K=2,\, b=0.01,\, T=0.001$).

A simple way to modify this system is to spring-load the rotor, thus adding a linear restoring force $ -k\,\theta $ in eq.\ref{eq:KR}, with $ k =$ spring constant (fig.\ref{fig:4sys}b). This system is known as Zaslavsky Web Map, and has been studied in the context of anomalous diffusion and quantum localization, with recent uses in cryptography \cite{zaslavsky2005hamiltonian, Mohammed2022ImageCF}. This map is now characterized by 2 parameters, $ K $ and $ f $ -- the natural frequency of the oscillator (so $ k=(2\pi f)^2 $, and we take kick period to be the unit of time). As before, while $ K $ controls the transition to chaos, $ f $ determines the symmetry of the phase-space structures. So for $ f=1/4 $ or $1/6$, we get the entire phase-space tiled with square or hexagonal lattice, respectively, of islands of regularity, making a sort of ``web'' (hence the name). It is perhaps surprising that islands of regularity are still found for arbitrarily irrational values of $ f $, and take various beautifully unexpected shapes (for sufficiently low $ K $) \cite{zaslavsky1992weak}. This illustrates the robustness of mixed chaos to complex conditions, and that regular dynamics may adapt to become quite complex themselves in such cases -- thus leading to rich dynamically ordered behaviors. As before, across all these regimes, our conjecture $ \ref{conj:1} $ was seen to hold in this system as well (for fig.\ref{fig:4sys}b, we used $ K=3, f=1/6, b=0.01, T=0.001$).

To illustrate the conjecture beyond kicked-rotor-like or other discretely driven systems, we show it for the Duffing oscillator (fig.\ref{fig:4sys}c):
\begin{align}
    &\dot{x} = v \nonumber\\
    &\dot{v} = -x^3 + x + F \,\sin(\omega\, t) - b\, v+ \sqrt{2\,b\,T} \; \xi(t)
\end{align}
This is similar to a driven harmonic oscillator, except we replace the quadratic potential, with the bistable $U(x) = x^4/4 - x^2/2 $ (fig.\ref{fig:4sys}c-i). Simulating this system is slower than the kicked systems, and requires full numerical forward-integration using sufficiently small time-steps $dt=0.01$. Being a simple non-linear generalization of the harmonic oscillator, while exhibiting a rich phenomenology including mixed chaos, strange attractors (above a critical value of $b$), stochastic resonance, and hysteresis, this system has been a key bridge between analytically tractable theory and key real-world complex behaviors\cite{kovacic2011duffing, coccolo2021delay, coccolo2024fractional}. The trajectories shown use the widely studied regime $F=0.3,\; \omega=1.2$, visualizing the stroboscopic Poincare map at $\omega\, t = 0 $ (mod $2\pi$). When coupling to a thermal bath via $b=0.01,\; T=0.01$, we again see emergent order through collapse to the regular island. Note that in the Hamiltonian trajectories fig.\ref{fig:4sys}(c-iii), we see trajectories outside the central chaotic region that seem like they could be regular -- however plot (c-ii) reveals that these still have a positive local Lyapunov exponent, unlike the truly integrable island of regularity in the middle. 

Another important and interesting example verifying the applicability of our conjecture \ref{conj:1} is the bouncing ball system, well-known as a simple experimental demo of rich chaotic dynamics \cite{Barroso2009BouncingBP, Naylor_Sanchez_Swift_2002}, with recent applications like energy-harvesting \cite{Huang2021ResearchOD}. Here an elastic ball is bouncing vertically (1D motion) in a uniform gravitational field on a sinusoidally oscillating plate:
\begin{align}
&Y_{plate}(t) = A\, \sin(\omega \, t) \nonumber\\
&\dot{y} = v \nonumber\\
&\dot{v} = -g - b \,v + \sqrt{2\,b\,T}\;\xi\\
&v \to -v + 2\,\dot{Y}_{plate} \quad \text{at bounce} \nonumber
\end{align}
Note that the thermal bath coupling here can be added via air-resistance (as in these equations) or via a coefficient of restitution and noise at each bounce. For small bath couplings, both methods produce the same results. 
In fig.\ref{fig:4sys}d, we plot these dynamics stroboscopically at the point of bounce (so, unlike other systems, \emph{not} at equal time-steps), showing the ball's post-bounce speed and plate's phase in the oscillation cycle at that moment. This is done for convenience, as the fastest way to simulate this system is at every bounce to numerically search for the time of the next bounce. It has previously been observed that addition of dissipation (like air resistance or energy loss at bounce) leads to regular dynamics, which are also ``quantized'' into discrete energy levels \cite{Jiang_Zhou_2021}, \cite{Naylor_Sanchez_Swift_2002}. Our conjecture explains this phenomenon, showing that such quantization merely reflects the phase-space structure of the non-dissipative Hamiltonian system, just as we saw with the kicked rotor earlier (fig.\ref{fig:4sys}d -- where the plate oscillates at 1 $ Hz $, with 0.1 $ m $ amplitude, $g =9.8\, m/s^2$, and bath coupling is introduced via $b=0.03$, $T=0.01$ or via coefficient of restitution = 0.98).

An interesting practical application we hypothesize for this result is in understanding and controlling the formation of corrugation, or ``washboard effect,'' on dirt roads.
In the reference frame of a car's wheel, we suggest that we could model it as a ball bouncing on the oscillating surface of the passing road. Then the spontaneous selection of the bouncing regular states we see in fig.\ref{fig:4sys}d would lead to a formation of entrenched corrugation. If so, these could therefore be disrupted by driving the system back into chaos, e.g., by reducing dissipation or introducing additional disorder -- perhaps in the road surface material, such as by using a gravel mixture with large-scale heterogeneity. Such interventions may not be obvious in the usual way of modeling these corrugations, which is to view them as a linearized dynamical instability or pattern formation \cite{BOTH2001Washboard}. This serves as a good example of a scenario where chaos may be preferable to the dynamical order arising from bath coupling.

Finally, we check that conjecture \ref{conj:1} also holds for systems with many degrees of freedom, pointing to its relevance for many-body self-organization. We consider a natural generalization of the web map which we term ``Kicked Harmonic Net'' (KHN) (fig.\ref{fig:4sys}e), which as far as we know has not been studied before. For $ N $ kicked rotors, in addition to pinning each one by a spring to $ \theta=0 $ as in the web map, we also connect them among each other in a harmonic oscillator network, thus giving the dynamics:
\begin{align}
\begin{pmatrix}
\v{\dot{\theta}}\\
\v{\dot{v}}
\end{pmatrix} = -
\overbrace{\begin{pmatrix}
0 & -I\\
k\,I + \Lambda & b\, I
\end{pmatrix}}^{W}\,
\begin{pmatrix} \v{\theta}\\\v{v} \end{pmatrix} 
\;\;-\; \begin{pmatrix} 0\\ K\,\sin(\v{\theta})\;\end{pmatrix}\delta(t-n) + \sqrt{2\,b\,T}\;\begin{pmatrix}0\\\v{\xi} \end{pmatrix} \label{eq:KHN}
\end{align}
Here $ \v{\theta} $ and $ \v{v} $ are now vectors of the $ N $ rotor coordinates, $ I $ is the identity matrix, and $ \Lambda $ is the connectivity graph Laplacian. This way, despite the vast complexity and flexibility of this system, it is still very fast to simulate as we have steps of linear evolution given by the matrix $ W $ followed by non-linear kicks. 
Note that kick amplitude and phase are here taken to be identical for all rotors.

The parameter space here is clearly too vast to survey fully, and so for a representative example, in fig.\ref{fig:4sys}e we took a randomly connected network of 16 rotors, where an edge was included in the network with probability 0.1 for every node pair (Erdős–Rényi model -- specific connectivity shown in fig.\ref{fig:4sys}(e-i), and parameters $ K=0.4, \,k=(\pi/5)^2,\, b=0.005,\, T=0.001$, with $ \Lambda $ scaled such that its mean eigenvalue is 2 $\Rightarrow \Lambda_{edge}\approx-0.84$). Since the phase space is 32-dimensional, we plot the $(\theta_1,v_1)$ phase space for just one of the oscillators. Note that here, the islands of regularity are so highly fine-tuned that no randomly sampled initial condition lands in them. 
Nonetheless, coupling to the bath reveals the presence of intricate set of ordered or partially ordered attractors -- fig.\ref{fig:4sys}(e-iv). To check that these attractors correspond to islands of regularity of the Hamiltonian system, we measure the local Lyapunov exponents at a set of points sampled from the attractors, as compared to points sampled randomly -- plot (e-ii) (note that we used a black background for the plot to highlight that unlike in other plots in column (ii), the phase space here is not sampled exhaustively). Similarly, in (e-iii), we ran Hamiltonian dynamics from set of IC combining random points with points from the attractors -- interestingly, only 2 of the 4 attractors in (e-iv) persisted under Hamiltonian evolution. 

It's important to note that unlike the other plots in column (iv), the dynamical order in the attractors here is not necessarily fully deterministic. This might at first seem contradictory to our conjecture \ref{conj:1}, for which islands of regularity were seen as integrable, and thus deterministic, dynamics. However, this behavior makes sense in the context of many-body self-organization such as the flocking of birds, where partial order is possible, with some birds forming a cohesive flock, while others are still moving independently and chaotically. In this scenario, it would be unwise to restrict our attention only to global order. Similarly, we can imagine a KHN where the springs pinning each rotor to $ \theta=0 $ are dominant compared to the ones among different rotors: $k \gg \Lambda$ (see eq.\ref{eq:KHN}). In the limit $\Lambda=0$ the rotors are independent, and clearly some can be dynamically ordered, while others are still chaotic, yielding chaotic motion on a lower dimensional submanifold of the full 32-dim. phase space. In this case, to find precisely the degree of partial order we could measure the fractal dimension of the trajectory. As we gradually restore the connectivity $\Lambda>0$, the different rotors' attractors start to interact. For large $\Lambda$ as in our example here, the dynamics no longer decompose by individual rotors, but some non-obvious partial order may still emerge -- and does emerge, as we see here. To quantify this, we measure the fractal dimension of each attractor (using the ``correlation dimension''), as labeled under plot (e-iv). 


While the rich phenomenology of KHN in its vast parameter space remains largely unexplored, our conjecture \ref{conj:1} continuous to hold in all the cases we tested. We observed all kinds of different complex attractor structures arise from weak coupling to thermal bath, including complex 1D knots, some with periods of hundreds of kicks before they repeat, 2D curved manifolds with interesting topologies, and higher dimensional attractors more difficult to visualize. This qualitatively continues to hold whether we change values of parameters like $K$ or $k$, number of rotors in the network, or the network topology (from random, to lattices, to small-world, etc). We invite further exploration of this system and its emergent behaviors in future work. 

Another interesting observation about this system is that it maps closely to Recurrent Neural Network architecture, where a weight matrix $ \exp{-W} $ is being iteratively applied to the state vector $ (\v{\theta},\v{v}) $ between nonlinear activation function given by the kicks (see eq.\ref{eq:KHN}). Providing input via time-varying kick strengths $ \v{K}(t) $ or via the initial conditions $ (\v{\theta}_0,\v{v}_0) $, one forward-pass thus corresponds to system evolution, while learning -- to changing the connectivity $\Lambda$. Emergence of a discrete set of dynamically ordered states implies an inherent tendency for compression and classification of the input signal in this architecture, even before any $ W $ learning. This can give more robust outputs and possibly better resistance to adversarial attacks, controlled by the bath coupling $ b $. At the same time, KHN may be a physically-realizable learning architecture, if we appropriately instantiate the dynamics of spring strengths on a slow timescale (see \cite{Chvykov_England_2018}). This builds on the recent interest to leverage self-organization in machine learning \cite{Vargas_Foong_Zhang_2023}.

\section{Transient} \label{sec:time}
\begin{figure}
\includegraphics[width=0.8\textwidth]{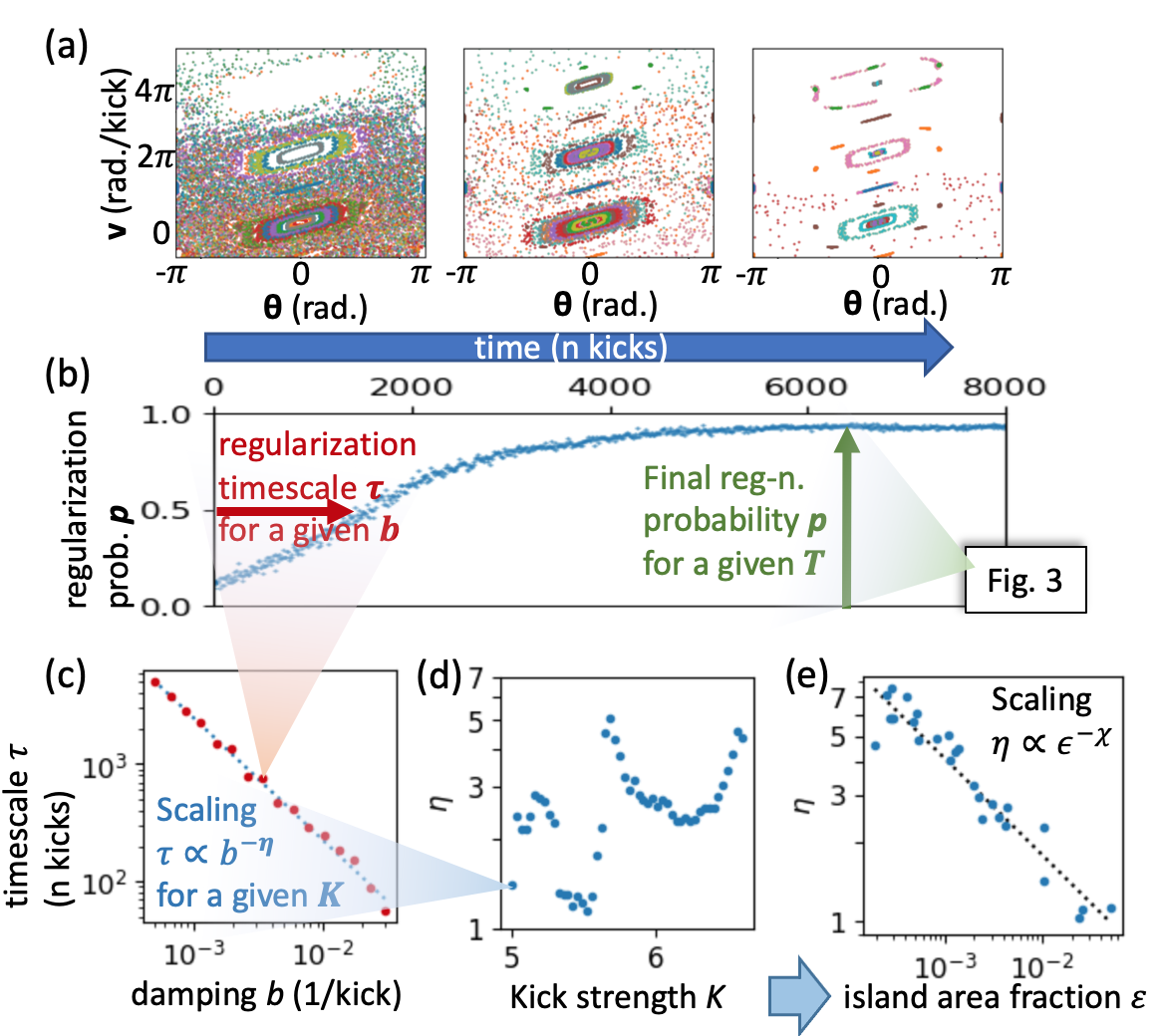}
\caption{Regularization timescale $ \tau $ and its dependence on damping $ b $. (a) kicked rotor relaxation dynamics (100 timesteps per frame at 3 points in evolution, with $K=2, b=0.002$); (b) corresponding relaxation for the fraction of random initial conditions that have regularized -- defining the timescale $ \tau $; (c) power-law dependence of $ \tau $ on $b $, which defines exponent $ \eta $; (d) $ \eta $ vs. kick strength $K$ shows no simple pattern, but falls neatly along straight line when plotted against $ \epsilon $ -- defined as the volume fraction occupied by islands of regularity in phase space (small for larger kick strength $ K $). This defines another scaling exponent $ \chi $, here $=0.4$.}
\label{fig:time}
\end{figure}

While conjecture \ref{conj:1} is powerful in its own right, its practical usefulness depends not only on the steady-state properties, but also on the relaxation timescale.
We mentioned that emergence of order should happen for arbitrarily small $ b $, but will take correspondingly longer. We sought to quantify this dependence with numerical experiments, focusing on the simple kicked rotor as it's the easiest to work with numerically.

To measure the regularization timescale we first need a way of detecting when a configuration is chaotic and when it's regular. For this, we used two methods. The simpler one is simply to count how many unique configuration were visited out of the past 20 steps, and if it is fewer than 16, then we roughly know that this is not chaos. This ``counting'' method will, however, fail to identify regular motion at non-zero temperature, as well as regular precession, such as in \ref{fig:4sys}(e-iv, red attractor). Therefore, for our second method, we directly measure local Lyapunov exponents by slightly perturbing the configuration, evolving both versions for 10 steps, finding a linear fit to log separation, and thresholding on the resulting slope. This method can cover the failure modes of the first one, but can get confused for weak chaos -- such as just outside regular islands. Therefore we use one or the other method depending on the specific context -- and verify that the two agree when both are applicable.

In fig.\ref{fig:time}b we see the resulting curve showing the probability of regularization over time at finite temperature. It is characterized by two parameters -- the regularization timescale $ \tau $, which we study here, and the steady-state probability of dynamical order $ p $, which we consider in the next section and fig.\ref{fig:T}. To recover the Hamiltonian regime, we expect that $ \tau $ should diverge as damping $ b \to 0 $ -- and in fig.\ref{fig:time}c we empirically show that this divergence is a power-law $ \tau \propto b^{-\eta} $. While the power-law behavior is universal across system parameters and the different systems we tried here, the scaling exponent $ \eta $ itself is not. In fig.\ref{fig:time}d, we thus show how $\eta$ varies with the kick strength $ K $. We see that this dependence is complex and non-monotonic. At the same time, $K$ is a parameter specific to the kicked rotor, while to find a general law, we want to use some universally relevant feature of the system. 
We hypothesize that $ \eta $ is primarily sensitive to the relative size of the islands of regularity $ \epsilon $ -- i.e., the ``degree of fine-tuning'' required for dynamical order. We define $ \epsilon $ as the fraction of randomly sampled initial conditions whose evolution is regular (regularity here measured by the ``counting'' method above). This way, for a set of kick strengths $ K $, we can measure both $ \eta $ and $ \epsilon $, and plot them against one another -- fig.\ref{fig:time}e. We see plotting $\eta$ against $\epsilon$ instead of $K$ neatly collapses all measurements onto one line. Although $ \eta $ does not vary over enough decades to see if this dependence is a power-law, at least locally it fits $ \eta \propto \epsilon^{-\chi} $ with $ \chi= 0.4$.

These results are important for several reasons. First, for regular inertial damped motion $ \dot{v}=-b\,v $, the relaxation timescale $ \tau = 1/b $, and so $ \eta=1 $. As we see in fig.\ref{fig:time}e, this value is approached when the islands of regularity become ``macroscopic,'' taking up a substantial fraction of the phase space ($\epsilon \gtrsim 0.05$). In this regime, regularization is ``trivial'' in the sense that it is just a matter of damping out extra velocity to relax to the nearest attractor -- since little fine-tuning is needed. Away from this limit, $ \eta >1 $ indicates non-trivial regularization dynamics. Such anomalous behavior can arise due to the ``stickiness'' of chaotic trajectories near regular islands, possible anomalous diffusion dynamics in the chaotic regions, or due to some kind of a ``search'' for the rare fine-tuned regular states. 

Second, since the relation $ \eta(\epsilon) $ in fig.\ref{fig:time}e does not refer to any system-specific quantities, it could be universal. We were able to confirm that it roughly holds for the web map (fig.\ref{fig:4sys}b), but could not reproduce it for kicked harmonic net (fig.\ref{fig:4sys}e). There, in addition to the heavier computation required, the high-dimensional phase space makes measuring $ \epsilon $ variations harder. Preliminary experiments showed that $ \eta \approx 5 $ seems to be relatively stable across parameters and even different number $ N $ of rotors. 

Such modification of $ \eta(\epsilon) $ relation in many-body systems could be attributed to the possibility of partial order, allowing a more gradual approach to regularity (see \cite{Simon_1962}). 
The simplest example to understand this is if instead of seeing fig.\ref{fig:4sys}b as $N$ different initializations of one kicked rotor, we saw it as describing a KHN with $N$ rotors and vanishing connectivity $\Lambda \to 0$. If we identify dynamical order by thresholding on Lyapunov exponent as we do now, then for this $2N$-dimensional system, this threshold would be crossed when enough of the rotors reach regular behavior -- i.e., when the regularization probability $p$ in fig.\ref{fig:4sys}b goes above some threshold. For large $N$ this will happen almost deterministically due to self-averaging, giving an apparently sharp onset of order with timescale $\tau \propto b^{-\eta}$ same as for a single rotor, independent of $N$. In contrast, the island area fraction in this $2N$-dimensional space will be $\epsilon_{N} = \epsilon_1^{\;N}$. So if we compare such ensembles with different $N$, we might have varying $\epsilon_N$, but constant $\eta$ -- similar to our preliminary observation for KHN. 
While this gives a reason for why $\eta$ might vary differently in a many-body system, it does not explain why $\eta$ should be constant across parameter regimes. This observation might indicate some many-body universality, where dynamics are dominated by collective modes arising from strong chaos and self-averaging, and insensitive to details of individual island structure.
These ideas needs further development, and we suggest that rather than being the full picture, our conjecture may be a fundamental building block with which we can build a more general theory of many-body self-organization.

Finally, this relation $ \eta(\epsilon) $ is important as it quantifies the usefulness of our main conjecture $ \ref{conj:1} $ in practically interesting cases, which may have strong fine-tuning (small $\epsilon$), such as for many-body self-organization. So if some scaling like $ \eta \propto \epsilon^{-0.4} $ in fig.\ref{fig:time}e was general, that would indicate that for extreme fine-tuning, this mechanism will have a very large $ \eta $, basically corresponding to random search of globally ordered states, which would predict unrealistically long self-organization timescales unless $ b $ is large. This way, the fact that we see $ \eta(\epsilon) $ being modified in our many-body example is promising evidence that the mechanism of conjecture \ref{conj:1} may be on the right track to explain cases of interest. 


\section{Noise} \label{sec:T}
\begin{figure}
\includegraphics[width=0.65\textwidth]{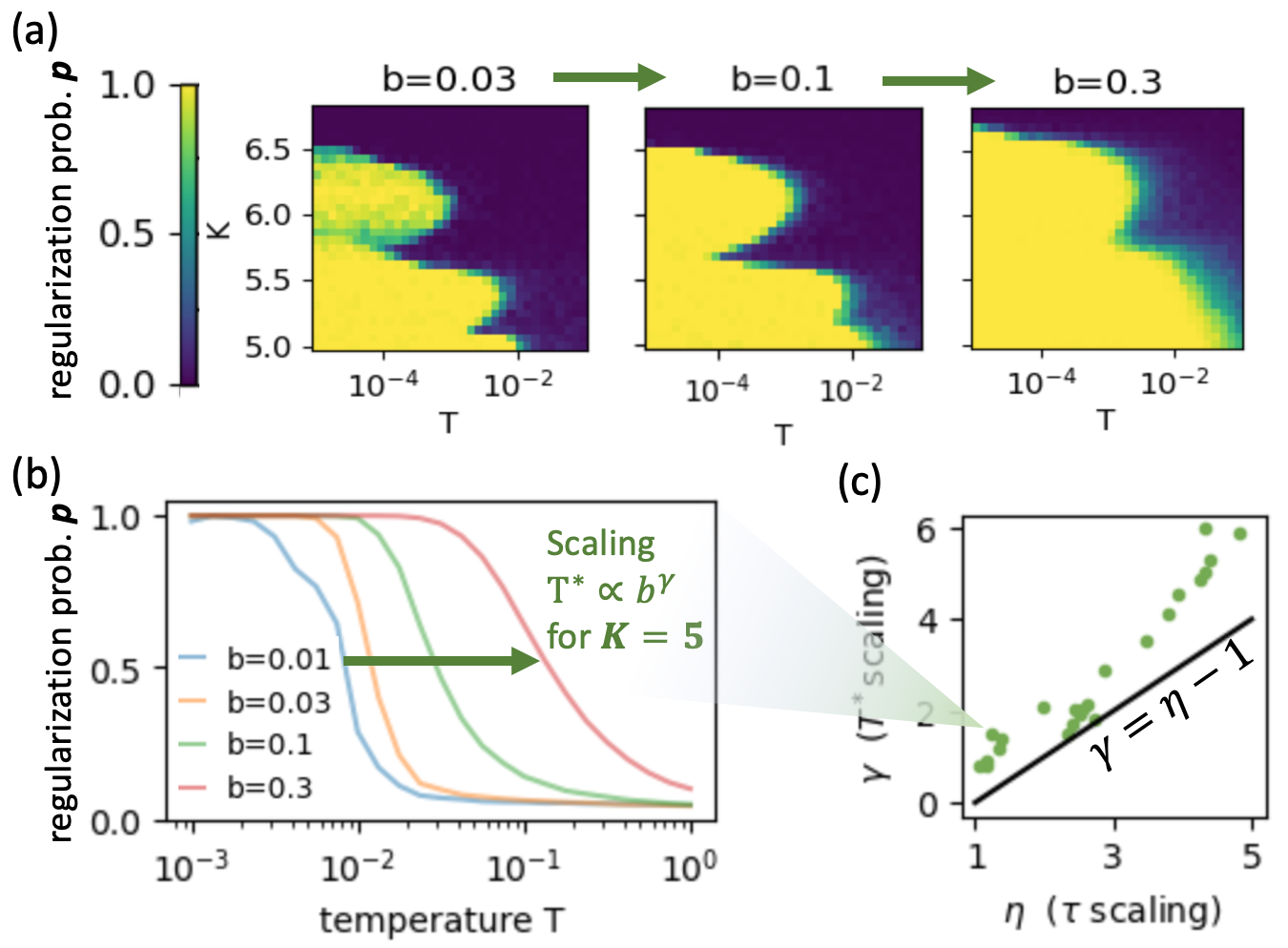}
\caption{Effects of thermal noise (on kicked rotor). (a) $K-T$ phase diagram at 3 values of $b$, color showing self-organization probability $ p $ at steady-state (see fig.\ref{fig:time}b). (b) phase boundary movement with $b$: the transition temperature $ T^* $ is observed to depend on $ b $ as power-law, defining scaling exponent $ \gamma $. (c) $ \gamma $ depends on system parameter $K$, and can be related to $ \eta $ (recall from fig.\ref{fig:time}d, $ \eta $ is large when $ K $ is). Line shows analytical prediction $ \gamma = \eta-1 $ -- suggesting an underlying fluctuation-dissipation relation (in equilibrium systems, Einstein relation gives $ \gamma=0 $ and $ \eta=1 $). $b$ is given in units of $1/kick$, and $T$ -- $(rad/kick)^2$.}
\label{fig:T}
\end{figure}

We said that conjecture \ref{conj:1} requires a low-temperature bath, hence we now want to quantify how raising the temperature breaks down emergent order.
Fig.\ref{fig:T}a shows how dynamical order is destroyed at high thermal temperatures, giving a ``phase diagram'' in the $K-T$ space (again we focus on kicked rotor here for simplicity). While it is tempting to think of this as an order-disorder phase transition, we must be cautious to remember that thermodynamic phase transitions are defined in terms of order parameters capturing the collective behavior of many degrees of freedom ($N\to\infty$), while here we are talking about a single kicked rotor dynamics. Nonetheless we can still loosely think of the two regimes as phases if we think of trajectories as a 1D chains along the time-dimension, and so the long-time average of the steady-state becomes our $N\to\infty$ limit. In that case, all the familiar properties of phase transitions can be studied, such as diverging correlation length, scaling exponents, etc.

Next, we check how the transition temperature $ T^* $ depends on damping $ b $ (fig.\ref{fig:T}a, b). Recall that for equilibrium systems in an energy landscape, the steady-state probability is independent of $ b $, which only controls the transients. Mathematically, this is ensured by the Einstein relation, which sets the noise amplitude to $ \sqrt{2\,b\,T} $ (see eq.\ref{eq:KR}). Thus the observation that here $ T^* $ depends on $ b $ at all is already interesting, indicating that we cannot simply view islands of regularity as acting effectively like energy wells. Moreover, we find approximately a power-law relation $ T^* \propto b^\gamma $, with $ \gamma $ -- a system-specific scaling exponent (fig.\ref{fig:T}b).

Interestingly, it seems that $ T^* $ indeed becomes independent of $ b $ ($ \gamma \approx 0 $) in the same regime where $ \tau \propto 1/b$ ($ \eta=1 $) -- i.e., where the relaxation dynamics are diffusion-like. Since Einstein relation is derived precisely by balancing transient relaxation with stochastic fluctuations in the steady-state Boltzmann distribution, and yields $\gamma=0,\, \eta=1$, we suspect that we can generalize this derivation to relate exponents $\gamma$ and $\eta$ in a nonequilibrium fluctuation-dissipation relation here. 

A simple way to model this relation is to approximate the system as a 2-state Markov process and consider the transition rates $ R $ between chaotic $ c $ and regular $ r $ dynamics. Such a simple model is motivated by realizing that in the undamped Hamiltonian system, islands of regularity $r$ and sea of chaos $c$ are isolated from each other, and so for small bath coupling $b$, we get perturbatively slow transition rates $\sim O(b)$\cite{jousseph2016weak}. This means that we can approximate the dynamics internal to $c$ and $r$ to erase memory of IC faster than the rare transitions among $r$ and $c$ can happen, justifying the Markov assumption. Then, in analogy to equilibrium phase transitions, which arise from competition between the concentrating effects of dissipation and the disordering effects of entropy, here $R_{c\to r}$ captures the former and $R_{r\to c}$ -- the latter. Our measurements of $ \tau $ in fig.\ref{fig:time} empirically give that $ R_{c\to r} \propto b^{\eta}$. To get $ R_{r\to c} $, we approximate the dynamics inside islands of regularity to be freely diffusing along $v$-dimension. This gives the exit rate in terms of the first-passage time of a diffusing particle in 1D to leave a region of size $ a $ -- which is known to be $ \propto a^2/D $. Since our phase space is 2D, we estimate $a^2 \propto \epsilon$, thus giving $R_{r\to c} \propto D/\epsilon$. From Einstein relation of the thermal bath, which drives the diffusion here, we get the diffusion coefficient $ D = b\,T $, and so $ R_{r\to c} \propto b\,T/\epsilon$. In the steady-state $ p_c \, R_{c\to r} = p_r\, R_{r\to c} $, from which we get that $ p_c = p_r$ at temperature $ T^* \propto \epsilon\,b^{\eta-1} $, thus predicting that $\gamma = \eta-1$.
Fig.\ref{fig:T}c shows that this roughly agrees with the data.

Being a discrete analogue of the derivation of Einstein relation, where $b^{\eta}$ played the role of motility $\mu$, while diffusivity $D$ was given by the usual expression here $b\,T$, we thus get that 
\begin{align} \label{eq:genE}
    D/\mu = b^{1-\eta}\, T
\end{align}-- a generalized Einstein relation, which reduces to the usual one for $\eta=1$. 

This simple model is clearly not the full story, but has the potential to be general. We were able to verify this relation also for the web map system, but not for the kicked harmonic net. There, the values of $\eta \approx 5,\, \gamma \approx 1$ seem to robustly persist across parameter regimes and different numbers of rotors -- even though the above derivation would still expect $\gamma = \eta-1$, regardless of the phase space dimensionality. This discrepancy makes sense when we remember that in the many-body case, the attractors we observed were only partially ordered, and therefore not deterministic (see fig.\ref{fig:4sys}e-iv). This means that the diffusivity inside the nearly-regular islands is not given purely by the thermal bath $D=b\,T$ as before, but can experience ``noise amplification'' when chaos stretches small thermal fluctuations. This modifies the exit rates $R_{r\to c} \propto D/\epsilon$, and hence the $\gamma \leftrightarrow \eta$ relation. To further refine the above model, one could also include a linear restoring ``force" $-b\,v$ in the first-passage time calculation, and relax the Markov assumption -- an undertaking we leave to future work.

\section{Conclusions}

In this work we used five different \emph{in silico} dynamical systems of varied structure and dimension to test the conjecture that Hamiltonian dynamics exhibiting mixed chaos will settle into islands of regularity upon coupling to a thermal bath at sufficiently low temperature. The practical power of this conjecture comes from showing that if a system is at all capable of regular motion, then cooling it by coupling to a cold thermal bath will tend to induce that regular motion regardless of initial conditions. Note that this means that cooling (or damping) does not generally reduce motion, but rather stabilizes one of the possible ordered dynamical patterns. Finally this conjecture allows making quantitative predictions about stability, relaxation times, and phase diagrams of the dynamically ordered behavior based on the knowledge of the phase space structure of the original Hamiltonian system. This can allow new ways to control dynamical systems by engineering the regular states and modulating their relative noisiness (similar to \cite{Chvykov_Berrueta2021} for example).
While we do not present a formal proof of conjecture \ref{conj:1}, leaving it for future work, we can now give some physical motivation for it. 


Our conjecture represents a bridge between two very different perspectives on emergence of dynamical order. In the thermodynamic perspective, the emphasis is on energy dissipation and lowering temperature as drivers of entropy reduction and emergent order (e.g., eq.\ref{eq:neqW}). In dynamical system theory, the focus is on damping, which allows contracting phase-space volumes, erasing memory of initial conditions, and turning small local Lyapunov exponents negative. But temperature and damping are two aspects of the same molecular dynamics of a thermal bath. Einstein's relation first established this connection between the thermodynamic and dynamical aspects of fluids. This way, while dissipative drag in undriven systems contracts phase-space volumes, turning all local energy minima into zero-dimensional dynamical attractors, thermal fluctuations add stochastic exploration of the phase space, together realizing the entropy-maximizing Boltzmann distribution. Here we want to generalize this insight for nonequilibrium systems (see eq.\ref{eq:genE}).


The analogy in driven Hamiltonian dynamics exhibiting mixed chaos arises because the dynamical counterpart of energy minima turns out to be the islands of regularity in the undamped driven system's phase space -- which similarly turn into absorbing attractors as soon as we add drag. One way to understand this is in the context of ``low rattling'' theory \cite{Chvykov_Berrueta2021, calvert2024local}: Initial conditions in the chaotic region execute a diffusive search of phase space until, at random, they enter a region of regularity. Once in such a region, and if temperature is low, the diffusive search shuts down due to the drop in chaotic motion, and density thereby accumulates around closed, ordered trajectories. This contraction of phase space density into lower sub-volumes is permitted by the frictional drag's violation of Liouville's theorem, which originally had to be obeyed by Hamiltonian dynamics in the absence of the heat bath. The key assumption of low rattling required for this argument is that steady-state probability of a state is controlled predominantly by that state's exit rate -- see \cite{calvert2024local} for a general study of this assumption.


This way, even for very small damping $ b $ we still expect self-organization in the steady-state, though the relaxation time $\tau $ can be long: we found that $\tau$ scales as power law $ \propto b^{-\eta} $ (fig.\ref{fig:time}). The scaling exponent $ \eta $ depends on system parameters, and may depend on the degree of fine-tuning required for dynamical order $\epsilon$ (defined as volume fraction of regular islands in the original Hamiltonian system) as $ \eta \propto \epsilon^{-\chi}$. We saw that in many-body systems, where the possibility of partial order makes self-organization more gradual, $\eta$ can be less sensitive to $\epsilon$.
Furthermore, as we vary temperature, our theory predicts a smooth breakdown of regularity, characterized by transition temperature $ T^* $ (fig.\ref{fig:T}). We find that $ T^*$ depends on $ b $, indicating the nonequilibrium nature of the phenomenon, and that $T^* \propto b^\gamma $. At equilibrium, $\eta=1,\,\gamma=0$. A simple model based on above understanding suggests a more general fluctuation-dissipation-type relation $ \gamma=\eta-1 $, which we validate experimentally. Interestingly, this relation is modified in systems with more degrees of freedom, indicating that other mechanics may be taking over, such as modification of free diffusivity inside the more complex ordered regions. 
Because of the broad implications this has for understanding nonequilibrium emergent order, we expect the conjecture on the effects of thermal bath coupling on mixed chaos presented here to stimulate further research across a range of disciplines.

\section*{Data availability statement}
The Python Jupyter notebook with code to simulate the presented systems and generate the datasets analyzed during the current study is available in the Github repository: \url{https://github.com/pchvykov/mixed_chaos_damped/}


\bibliography{mixed_chaos.bib}



\section*{Author contributions statement}

P.C. conceived of the main results, did the simulations and data analysis, under the guidance of and with regular feedback from J.E. All authors reviewed the manuscript. 

\section*{Additional information}

\textbf{Competing interests}: the authors declare no competing interests.

\end{document}